\documentstyle[prl,aps]{revtex}
\begin{document}

\title{Yang--Mills analogues of the Immirzi ambiguity}

\author{Rodolfo Gambini$^1$, Octavio Obreg\'on$^2$, Jorge Pullin$^3$}
\address{1. 
Instituto de F\'isica, Facultad de Ingenier\'{\i}a,
J. Herrera y Reissig 565, Montevideo, Uruguay}
\address{2.
Instituto de F\'{\i}sica, Universidad de Guanajuato,
PO Box E-143, C. P. 37150 \\ Le\'on, Guanajuato,
M\'exico}
\address{
3. Center for Gravitational Physics and Geometry, Department of
Physics,\\
The Pennsylvania State University, 
104 Davey Lab, University Park, PA 16802}

\maketitle
\begin{abstract}
We draw parallels between the recently introduced ``Immirzi
ambiguity'' of the Ashtekar-like formulation of canonical quantum
gravity and other ambiguities that appear in Yang--Mills theories,
like the $\theta$ ambiguity. We also discuss ambiguities in the
Maxwell case, and implication for the loop quantization of these
theories.
\end{abstract}
\vspace{-7cm} 
\begin{flushright}
\baselineskip=15pt
CGPG-98/1-2  \\
gr-qc/9801055\\
\end{flushright}
\vspace{5.5cm}

\pacs{4.60 Ds} 

\section{Introduction}

Several new insights into the canonical quantization of
general relativity have been acquired using Ashtekar-like
variables \cite{As}. Originally, this consisted in basing the theory on a
canonical pair formed by a set of (densitized) triads $\tilde{E}^a_i$,
and a (complex) $SU(2)$ connection $A_a^i \equiv \Gamma_a^i +i K_a^i$,
where $\Gamma_a^i$ is the spin connection compatible with the triads
and $K_a^i= K_{ab} E^{bi}$ where $K_{ab}$ is the extrinsic curvature.
In terms of these variables, the constraints of the theory became a
Yang--Mills- like Gauss law, plus expressions for the traditional
vector and Hamiltonian constraints,
\begin{eqnarray}
D_a \tilde{E}^a_i &=& 0 \label{gauss}\\ 
\tilde{E}^a_i F_{ab}^i &=& 0\label{diffeo}\\ 
\epsilon^{ijk} \tilde{E}^a_i \tilde{E}^b_j F_{ab}^k
&=& 0 \label{ham}. 
\end{eqnarray} 

It was first noted by Barbero \cite{Ba}, that a reasonably similar
structure could be achieved in terms of a one-parameter family of
variables. If one considers a connection of the form $A_a^i =
\Gamma_a^i + \beta K_a^i$, with $\beta$ an arbitrary complex number,
it can be shown that the vector and Gauss-law constraints retain
exactly the same form as (\ref{gauss},\ref{diffeo}), provided one
re-scales the triads by an overall $1/\beta$ factor. The form of the
Hamiltonian constraint changes. Immirzi \cite{Im} first noted that the
availability of this one-parameter family of connections led to
apparently puzzling results. Due to the complexity of the Hamiltonian
constraint (\ref{ham}), a significant portion of the work on canonical
quantum gravity has up to now concentrated on ``kinematics''. This
refers to the study of features that only depend on the structure of
the Gauss law and vector constraints
(\ref{gauss},\ref{diffeo}). Examples of this kind of work are the
quantization of area and volume \cite{RoSmvo,AsLe}.  These results
have direct impact on more attractive ``physical'' issues as the
recent attempts to compute black hole entropy in nonperturbative
quantum gravity \cite{Kr,Ro,AsBaCoKr}. What Immirzi noticed is that in
spite of the fact that different values of $\beta$ leave the
constraints (\ref{gauss},\ref{diffeo}) invariant, the spectra of
certain quantum operators depend on $\beta$. An example of this
property is the area operator, whose spectra in terms of spin network
states depends on an overall $\beta$ factor. Rovelli and Thiemann
\cite{RoTh} noted that the different conjugate pairs
$(\tilde{E}^a_i,A_a^i)$ constructed with different $\beta$ differed by
a canonical transformation. However, this canonical transformation was
not being unitarily implemented in the quantum theory. Thus, the
changes in the spectra of physical operators. The fact that the change
in spectra had direct impact in ``observable'' computations, like the
entropy of a black hole, motivates trying to understand better the
role that the $\beta$ parameter has in canonical quantum gravity.  The
purpose of this paper is to discuss this. We will note that the role
of the $\beta$ parameter in canonical quantum gravity is analogous in
various senses to that of the $\theta$ parameter that describes the
different sectors associated to the topological structure of large
gauge transformations in Yang--Mills theory. In particular we will
notice that loop representations appear only capture one such
``sector'' at a time.

The organization of this paper is as follows, in the next section we
discuss the Immirzi ambiguity, in section III we draw a parallel with
the $\theta$ ambiguity of Yang--Mills theories and in section IV we
study the case of Maxwell theory.

\section{The Immirzi ambiguity}

In the gravitational case, the Immirzi ambiguity arises as a canonical
transformation that is not implemented unitarily in the quantum theory
in terms of the loop representation. In such case one is using a basis
of states (diffeomorphism invariant functions of loops)
that is invariant under small gauge and diffeomorphism
transformations. If one writes Barbero's Hamiltonian in terms of loops
it would be  
$\beta$-dependent and the
physical quantities, such as the area, are also $\beta$-dependent.  To
emphasize the analogy with the Yang--Mills case, let us write the
action for general relativity in a Palatini form in terms of tetrads,
but also add to it a term that vanishes on-shell, as suggested in
\cite{Ho},
\begin{equation}
S = {1 \over 2} \int {\rm Tr}( \Sigma\wedge R) -{1 \over \beta }
{\rm Tr}( \Sigma\wedge {}^*R), 
\end{equation}
where $\Sigma = e \wedge e$, $e_a^I$ being a tetrad and $R_{ab}^{IJ}$
is  the curvature associated with the spin connection compatible with
the tetrad. ${}^*R_{ab}^{IJ} \equiv \epsilon^{IJ}_{KL} R_{ab}^{IJ}$.
It is well known that the added term vanishes on-shell, this was
the key idea that launched the original Ashtekar new variables (which are
obtained taking $\beta=i$), allowing to use a complex action to
describe a real theory without adding new equations since the
imaginary part of the action is topological in nature. If one performs
a canonical decomposition of this action, the canonically conjugate
pair is given by a densitized triad $\tilde{E}^{a}_i$, playing the
analogous role to the electric field in a Yang--Mills theory and a
connection $A_a^i = \Gamma_a^i + \beta K_a^i$.

The Gauss law and the vector constraint are not $\beta$ dependent
(strictly speaking, this means that one can always find linear
combinations of these constraints that are $\beta$-independent). This
is suggested at the level of the action by the fact that the action is
diffeomorphism and gauge independent for all values of $\beta$. The
Hamiltonian constraint, however, is $\beta$ dependent,
\begin{equation}
H= \epsilon^{ijk} \tilde{E}^a_i \tilde{E}^b_j F_{ab}^k
-2 (1+\beta^2) \tilde{E}^{[a}_i \tilde{E}^{b]}_j K_a^i K_b^j
=0. 
\end{equation}
where $K_a^i = (A_a^i -\Gamma_a^i)/\beta$ is related to the extrinsic
curvature.

The $\beta$ dependence of the Hamiltonian shows that the resulting 
physics of quantum gravity will be $\beta$-dependent in
general. Therefore, one could fix the value of the parameter $\beta$ 
``experimentally''. What is more surprising, is that physical
quantities that do not have to do with the Hamiltonian, also end up
being $\beta$ dependent. A typical example is the area operator. If
one considers a surface $S$ and computes the quantum operator in
the loop representation for the area of such surface one finds that in
the basis of spin networks the operator is given by 
\cite{AsBaCoKr},
\begin{equation}
\hat{A} |\Gamma> = 8 \pi \beta \ell_{\rm Planck}^2 \sum_p \sqrt{j_p (j_p+1)},
\end{equation}
where $j_p$ are the valences of the $p$ lines of the spin network that
cross the surface $S$.

This raises the question of what is the nature of these ambiguities
and if similar ambiguities are present in other theories. As we
mentioned in the introduction, these ambiguities correspond to
canonical transformations that are not being unitarily implemented in
the quantum theory. We also may add that the transformations preserve
the form of the ``kinematical'' constraints of the theory.  We will
see in the following sections that similar ambiguities may arise in
gauge theories.

\section{The $SU(2)$ Yang--Mills case}

Let us briefly recall the $\theta$ ambiguity in Yang--Mills theory
(for a more complete discussion see \cite{Ja,tH}). If one starts from
the Yang--Mills action, $S = {1 \over 4 g^2} {\rm Tr} [\int F\wedge
{}^*F]$ and performs a canonical formulation of the theory, one finds
that the quantum Gauss Law constraint ensures invariance of the
wavefunction under gauge transformations connected with the
identity. Wavefunctions in general are not invariant under large gauge
transformations, characterized by a winding number $n$. We denote by
$\hat{\Omega}_n$ the generator of large gauge transformations,
$\hat{\Omega}_n \Psi[A] = \Psi[g\cdot A\cdot g^{-1}+ g\partial
g^{-1}]$, where $g$ is the gauge transformation matrix for a gauge
transformation with winding number $n$. $\hat{\Omega}_n$ is a unitary
operator that commutes with the Hamiltonian of the theory. 

One can therefore construct a basis of common eigenstates of
$\hat{\Omega}_n$ and the Hamiltonian, labelled by the eigenvalues of
$\hat{\Omega}_n$,
\begin{eqnarray}
\hat{\Omega}_n \Psi_\theta[A] &=& \exp(i\theta n)\Psi_\theta[A]
\label{large}
\\
\hat{H} \Psi_\theta[A] &=& E_\theta \Psi_\theta[A].
\end{eqnarray}
We therefore see that the quantum theory contains an infinite number
of disjoint sectors labelled by the continuous angle $\theta$. If one
is working in the connection representation, as we have done up to
now, one is able to describe simultaneously all the disjoint sectors.
However, if one wishes to consider the loop representation, things are
different. Since the basis of Wilson loops is invariant under large
gauge transformations, it can only give rise to functions that are
invariant under large gauge transformations, or in terms of equation
(\ref{large}), to the sector $\theta=0$. That is, the loop
representation only captures one of the $\theta$ sectors of the
theory \cite{FoGa}.

If one now considers a new action for the theory, obtained
by adding the Pontryagin topological term to the ordinary Yang--Mills
action,
\begin{equation}
S = {1 \over 2 g^2} {\rm Tr} [\int F\wedge {}^*F] +
{\theta_0  \over 16 \pi^2} {\rm Tr} [\int F\wedge F], \label{acteta}
\end{equation}
the classical theory is unchanged since one added a total divergence
to the action. The added term only contributes a Chern--Simons type 
term evaluated on the boundary of the manifold, and is invariant under
gauge transformations connected with the identity, changing by an
integer value for large gauge transformations.

If one constructs a canonical formulation starting from action
(\ref{acteta}), the resulting electric field is related to that of the
original action by $E = E_{\rm orig}+ {\theta_0 g^2 \over 8\pi^2}
B_{\rm orig}$ where $B$ is the magnetic field. The resulting theory
has the same physical predictions as the one we considered before.
There is a relationship between the description of both given by
$\Psi[A] = \exp(i W[A] \theta_0) \Psi[A]_{\rm orig}$, where
$W[A]=(-1/16\pi^2) {\rm Tr}[ \int F\wedge A -2/3 A\wedge A \wedge A]$
is the integral of the Chern--Simons form. The new theory has the same
$\theta$ structure for the vacua as the one originally considered,
the $\theta$ angles being shifted by $\theta_0$, in the sense that,
\begin{eqnarray}
\Omega_n \Psi[A] &=& \exp\left( i (\theta-\theta_0) n \right) \Psi_\theta[A]
\label{variation}\\
\hat{H}_{\theta_0} \Psi_\theta[A] &=& E_\theta \Psi_\theta[A].
\end{eqnarray}

If we now consider the loop representation, the Hamiltonian of the
theory is $\theta_0$ dependent, and so are its eigenvalues. The loop
representation still captures a single $\theta$ sector of the theory,
but now for a different value, given by the parameter $\theta_0$.
Therefore, it is clear that the canonical transformation we just
introduced is not being unitarily implemented in the loop
representation, since the spectra of the Hamiltonian changes.

We see that there are clear parallels (and distinctions) between the
$\theta$ ambiguity and the Immirzi ambiguity. In both cases, one finds
physical quantities that depend on the ambiguity. The ambiguity is
``resolved experimentally''  when one considers the full dynamics of
the theory, since in both cases the Hamiltonians depend on the
parameters in question. In both cases the ambiguities correspond to
canonical transformations at a classical level. 

The main difference between both ambiguities is due to the extra term
in the action one adds in both cases is of a different nature. In the
$\theta$ ambiguity it is a total divergence. This allows a deeper
understanding of the $\theta$ sectors as related to the topological
structure of large gauge transformations, and the identification of
the corresponding $\theta$ sectors. Such understanding is lacking in
the case of the Immirzi ambiguity, which is generated by a term in the
action that vanishes on shell, but is not a total divergence.

\section{Maxwell theory}

It has been noticed by Corichi and Krasnov \cite{KrCo} that free
Maxwell theory has an Immirzi-like ambiguity consisting of rescaling
the electric field and vector potential by a constant $\epsilon$ in
such a way as to preserve the canonical commutation relations. If one
constructs a loop representation in terms of the connection
$A_a/\epsilon$ one can see that one can define an operator
representing the charge enclosed by a surface that works in an
analogous way as the area operator in quantum gravity. Its spectrum is
rescaled by $1/\epsilon$. Therefore there is a parallel with the
gravitational case, the charge playing the role of the area
observable. It is worthwhile noticing, however, that the above
ambiguity does not survive the coupling of the theory to matter. If
one adds electric charge, Gauss' law implies that one cannot re-scale
the electric field unless one changes the charge in the theory.
Another way of seeing this is to consider the theory coupled to
Fermions and build a loop representation. If one does so, requiring
that the holonomy with Fermions inserted at its ends be a gauge
invariant quantity uniquely fixes the $\epsilon$ parameter.  We would
therefore like to concentrate on other types of ambiguities in Maxwell
theory that would survive the inclusion of matter. The Immirzi
ambiguity in gravity does not change if one couples the theory to
matter. If one considers non-Fermionic matter, there is no
contribution to the (gravitational) Gauss law and the contributions to
the vector constraint do not involve the connection and therefore are
$\beta$-independent. For Fermions, there is a contribution to the
Gauss law proportional to $\psi^\dagger \psi$, but the gravitational
part of it is $\beta$ independent. For the vector constraint, the
gravitational part and the Fermionic piece are $\beta$ dependent, but
one can see that the portion depending on $\beta$ is proportional to
the Gauss law.

One can introduce ambiguities in Maxwell theory that survive the
inclusion of matter by considering $\theta$ ambiguities. The
discussion goes through very much as in the $SU(2)$ case, but with one
important difference: in $3+1$ dimensions there are no large gauge
transformations associated with the $U(1)$ group, so for all practical
purposes the ambiguity is not there. One can add to the action a
$\theta$ term, but physical quantities do not change their
spectra. Loop representations can be built and although their
appearance is different, one can see that they are unitarily related. 
This is accomplished by noticing that for the Abelian case one can
find an expression for the Chern--Simons factor in the loop
representation, built using the connection and loop derivatives
\cite{GaPubook}, since there is no problem with large gauge
transformations. 

One can construct an analog of the $\theta$ ambiguity for the Maxwell
theory in $1+1$ dimensions, and the situation is completely analogous
to the Yang--Mills case in higher dimensions, see \cite{GaMoUrVe} for
references.

There is a different type of ambiguity that arises in Maxwell theory.
This is slightly different from the theta ambiguity and has parallels
with the Immirzi case. This arises from the fact that one can
introduce more than one connection\footnote{For Maxwell theory one can
introduce more than one connection and also more than one electric
field. This is easily seen in the analogy with the harmonic oscillator
in the Bargmann representation, where one can take as canonical pairs
$(z,\bar{z})$ with $z=q+ip$, or $(q,z)$ or $(p,z)$, etc. Ashtekar and
Rovelli choose mixed variables for both the connection and the
electric field in their treatment of the Maxwell theory.}  for Maxwell
theory. This was first noticed by Ashtekar and Rovelli \cite{AsRo}, by
drawing an analogy with the Bargmann quantization. They considered as
canonical variables for Maxwell theory the positive frequency
connection,
\begin{equation}
{}^+A_a = {1 \over \sqrt{2}} (A_a^T(x) + i {1\over \Delta^{1/2}} E_T^a(x))
\end{equation}
with $T$ standing for transverse, and $\Delta$ is minus the
Laplacian operator. If we now consider a more general connection,
\begin{equation}
{}^\beta A_a = {1 \over \sqrt{2}} (A_a^T(x) + \beta {1\over
\Delta^{1/2}} E_T^a(x))
\end{equation}
we can construct a family of quantum theories. The transformation is
clearly a canonical transformation. Yet, if one goes to the loop
representation they are not necessarily implemented unitarily, as we 
will immediately see.

An interesting aspect that is worthwhile pointing out 
is that in this context certain values of $\beta$ are preferred
purely from mathematical considerations. If one considers $\beta$
real, and one tries to construct a loop representation, one ends up
with the same problems as the first attempts found
\cite{DiGaTr,GaPubook}. Namely, the Fock space wavefunctions are not
well implemented in the loop representation. This difficulty was
circumvented by Ashtekar and Rovelli by considering $\beta=i$. One can
see that the problem does not arise for $Im(\beta)$ nonzero. Clearly
these two representations cannot be unitarily connected.

It is worthwhile pointing out that this ambiguity survives the
inclusion of matter, it is perfectly possible to discuss Maxwell
theory coupled to Fermions in terms of these variables without fixing
the value of $\beta$.

\section{Conclusions}

In this paper we have pointed out that ambiguities similar to the one
Immirzi encountered in gravity exist in other theories, in particular
in Yang--Mills theory. This confirms what was pointed out by Rovelli
and Thiemann, in the sense that one ``needs two connections'' for
Immirzi-like ambiguities to arise. What we see is that through the
addition of $\theta$ terms one accomplish essentially the same by
having ``two electric fields'', and introducing a canonical
transformation that preserves the Gauss law constraint.  For Maxwell
theory, one can take advantage of the simplification in Gauss' law
that arises in the Abelian case to again introduce ``two electric
fields'' or ``two connections'' (or combinations thereof), and end up
with ambiguities. We see that for the Maxwell case the ambiguity can
be eliminated partially in the loop representation by requiring that
the Fock space structure be properly represented. It is worthwhile
considering if a similar selection based on purely mathematical
criteria might be present in the case of quantum gravity.

\acknowledgements

We wish to thank Abhay Ashtekar, Hugo Fort, Don Marolf, Giorgio
Immirzi and Thomas Thiemann for discussions.  This work was supported
in part by grants NSF-INT-9406269, NSF-INT-9722514, NSF-PHY-9423950,
research funds of the Pennsylvania State University, the Eberly Family
research fund at PSU and PSU's Office for Minority Faculty
development. JP acknowledges support of the Alfred P. Sloan foundation
through a fellowship. We acknowledge support of Conicyt (project 49),
PEDECIBA (Uruguay) and Conacyt (Mexico), through grant 3898P-E9608.

\end{document}